\documentclass[twocolappendix,numberedappendix]{emulateapj}
\usepackage[z,alpha]{tmsfnts}

\usepackage{epstopdf} 

\newcommand{\msol}{\hbox{${\rm M}_\odot$}}

\newcommand{\LCDM}{$\Lambda$CDM }

\newcommand{\kms}{km\,s$^{-1}$}
\newcommand{\hinv}{h^{-1}}
\newcommand{\mpc}{\rm{Mpc}}
\newcommand{\hmpc}{\hinv\mpc}
\newcommand{\kpc}{\rm{kpc}}

\shorttitle{Mass of the Local Group}
\shortauthors{Gonz\'alez, Kravtsov \& Gnedin}

\begin{document}
\title{On the mass of the Local Group}

\author{Roberto E. Gonz\'alez\altaffilmark{1,2,5}}
\author{Andrey V. Kravtsov\altaffilmark{1,2,3}}
\author{Nickolay Y. Gnedin\altaffilmark{1,2,4}}
\altaffiltext{1}{Department of Astronomy \& Astrophysics, The University of Chicago, Chicago, IL 60637 USA}
\altaffiltext{2}{Kavli Institute for Cosmological Physics, The University of Chicago, Chicago, IL 60637 USA}
\altaffiltext{3}{Enrico Fermi Institute, The University of Chicago, Chicago, IL 60637 USA}
\altaffiltext{4}{Particle Astrophysics Center, Fermi National Accelerator Laboratory, Batavia, IL 60510 USA}
\altaffiltext{5}{Instituto de Astrof\'{i}sica, Pontificia Universidad Cat\'olica de Chile, Santiago, Chile {\tt regonzar@astro.puc.cl} }

\begin{abstract}

We use recent proper motion measurements of the tangential velocity of M31, along with its radial velocity and distance, to derive the likelihood of the sum of halo masses of the Milky Way and M31. This is done using a sample halo pairs in the Bolshoi cosmological simulation of $\Lambda$CDM cosmology selected 
to match properties and environment of the Local Group. The resulting likelihood gives 
estimate of the sum of masses of  $M_{\rm MW,200c}+M_{\rm M31,200c}=$ $2.40_{-1.05}^{+1.95}\times10^{12}\,M_{\odot}$ ($90\%$ confidence interval). This estimate is consistent with individual mass estimates for the Milky Way and M31 and is consistent, albeit somewhat on the low side, with the mass estimated using the timing argument. We show that although the timing argument is unbiased on average for all pairs, for pairs constrained to have radial and tangential velocities similar to that of the Local Group the argument overestimates the sum of masses by a factor of $1.6$. 
Using similar technique we estimate the total dark matter mass enclosed within $1$ \rm{Mpc} from the Local Group barycenter to be $M_{\rm LG}(r<1\, {\rm Mpc})=4.2_{-2.0}^{+3.4}\times10^{12}\,M_{\odot}$ ($90\%$ confidence interval).
\end{abstract}


\section{Introduction}

Understanding the connection between dark matter (DM) halos and galaxies they host is 
a key question in galaxy formation theory. Theoretical models of hierarchical structure
formation \citep{1978MNRAS.183..341W,fall_efstathiou80,blumenthal_etal84} envision dark matter halos to be the sites of galaxy
formation and this framework is supported by a variety of observations \citep[see, e.g., recent reviews by][]{frenk_white12,courteau_etal14}, such as galaxy rotation curves
\citep{1970ApJ...159..379R,roberts_rots73}, X-ray halos  \citep[][see \citealt{mathews_brighenti03} for a review]{forman_etal85,buote_canizares94,buote_etal02,humphrey_etal11,akos_etal13}, satellite kinematics \citep{zaritsky_etal93,zaritsky_etal97,zaritsky_white94,mckay_etal02,prada_etal03,conroy_etal07,klypin_prada09,more_etal11}, and weak lensing measurements \citep[e.g.,][]{mandelbaum_etal06,van_uitert_etal11,velander_etal13,hudson_etal13}.

The Local Group (LG hereafter) played an important role in establishing existence of extended massive halos around galaxies. Indeed, the first flat rotation curve  was measured for M31 \citep{babcock39} and the mass estimate for the LG by 
\citet{1959ApJ...130..705K} was one of the very first compelling indications for existence of massive dark matter halos. The elegant argument in the latter study relied on the assumption that LG can be approximated by two point masses on a radial orbit on the first approach. Orbit integration backward in time, given the present day separation, velocity, and cosmological parameters, then constrains the mass of the system. This framework is now known as the timing argument (TA hereafter). Despite its simplicity and strong assumptions, the argument has withstood the test of time and new observations \citep{2008MNRAS.384.1459L,2008ApJ...678..187V,2012ApJ...753....8V}. Nevertheless, the masses of both the Milky Way and M31 are both uncertain to a factor of two \citep[e.g.,][and references therein]{boylan_kolchin_etal13}. Given the large uncertainties, the LG mass derived using the timing argument \citep{2008MNRAS.384.1459L,2012ApJ...753....8V} is generally consistent with  mass estimates derived using other methods \citep[][and references therein]{2002ApJ...573..597K,widrow_dubinsky05,2009MNRAS.393.1265K,2010MNRAS.406..264W}, but is on the high side of the measurement range. 

In this paper we present a different way to constrain the mass of the Local Group using approach similar to that used by \citet{2011ApJ...743...40B} to constrain the mass of the Milky Way. In this approach a set of observed properties of a system is used to estimate likelihood that a system in
simulation is a counterpart of this system. Distribution of the likelihood as a function of halo mass can then be used to estimate the mass of observed system. In this study we select a population of the LG analogues from the Bolshoi cosmological simulation of \LCDM cosmology \citep{2011ApJ...740..102K} and use observed properties of the MW and M31 to derive likelihood distribution for their combined mass.

We use several criteria to define the LG pair analogues in the cosmological simulation. In addition to distance and radial and tangential velocities, we also consider parameters characterizing environment, such as the distance to the nearest cluster, local large-scale density, and coldness of the local galaxy flow. The Local Group is known to reside in a region of rather low (``cold'') radial velocity dispersion of galaxies, $\sigma_{\rm H}<70$\kms \citep[][and references therein]{1975ApJ...196..313S,1997NewA....2...91G,2003AA...398..479K,2003ApJ...596...19K,2011MNRAS.415L..16A}, as compared to velocity dispersion around MW-sized halos in the \LCDM cosmology. This can be explained by the fact the LG is located in an average density environment \citep{2003ApJ...596...19K}. We will explore in more detail this density and velocity dispersion relation, and use it to impose additional constrains to our LG analogues.

The paper is organized as follows. In \S~\ref{sec:sim} we describe the simulation and halo catalogs, while in \S~\ref{sec:lgsel} we describe selection criteria for the LG analogues and different synthetic LG samples. We present our results for the likelihood distribution of the LG mass in \S~\ref{sec:like} and compare mass estimated using this method with previous estimates using the timing argument in \S~\ref{sec:ta}. We discuss our results and summarize conclusions in \S~\ref{sec:conc}. In this paper we use mass, $M_{200c}$, defined as  the mass within radius $R_{200c}$ enclosing the mean density of $200$ times the critical density at the redshift of analysis. 
For the Milky Way-sized halos, $M_{200c}$ is related to the commonly used virial mass definition defined using cosmology and redshift dependent overdensity \citet{1998ApJ...495...80B} as $M_{\rm vir}/M_{200c}\approx 1.2$.

\section{Simulations and halo catalogs}
\label{sec:sim}

To construct a sample of the Local Group analogues, we use halos from the Bolshoi simulation of $\Lambda$CDM cosmology: $\Omega_{\rm m}=1-\Omega_{\Lambda}=0.27$, $H_0=70\,\rm km/s/Mpc$, $\sigma_8=0.82$, $n_s=0.95$ \citep{2011ApJ...740..102K}, compatible with the constraints from the WMAP satellite \citep{hinshaw_etal13}. The simulation followed evolution of dark matter in a $250 \hmpc$ box with spatial resolution of $\approx 1h^{-1}$~kpc and mass resolution of $m_{\rm p}=1.35\times 10^8\ \rm M_{\odot}$. Halos are identified with the BDM algorithm  \citep{1997astro.ph.12217K}. The BDM algorithm is  a spherical overdensity halo finding algorithm and is designed to identify both host halos and subhalos. In this study, however, we will only use the host halos. 

The catalog of host halos is complete down to halos with maximum circular velocities of $\approx 50\ \rm km\,s^{-1}$, and we use only halos of larger mass to identify pairs of the MW-sized halos. 
To construct a sample of the MW-M31 pairs at $z\approx 0$, we use a series of simulation snapshots  at $z<0.1$ (i.e. in the last $\approx 1.3$ Gyr before present) spaced by $\approx 150-250$ Myr, similarly to the strategy adopted in \citet{2013ApJ...770...96G}. This is done because a particular configuration of MW and M31 is transient and would correspond to a relatively small number of systems at one snapshot. By using multiple snapshots we can increase the sample of systems in such configuration during a period of time in which secular cosmological evolution is small. For instance,the average mass growth of MW sized halos in the simulation since $z=0.1$ is only $1.2\%$.

\section{The sample of Local Group analogues}
\label{sec:lgsel}

The Local Group is dominated by the pair of the Milky Way and M31 and includes a 
number of smaller galaxies. Environment around the Local Group has density quite
close to the average density of the universe \citep{2003ApJ...596...19K,2005AJ....129..178K,karachentsev12}. In addition, the closest massive galaxy cluster, the Virgo Cluster, is $\approx 16.5\ \mpc$ away \citep{2007ApJ...655..144M}. 
It is not clear to what extent the environment of the Local Group shapes its properties and dynamics. Therefore, we include environmental criteria in our set of selection criteria. 

In order to identify the LG analogues,
we select pairs in relative isolation and in a wide range of masses from  $M_{200c}=5 \times 10^{10}$ \msol $ $ to $ 5 \times 10^{13}$ \msol. To avoid pairs in triplets or larger groups we define a quantitative isolation criterion using the force constraint $F_{i.\rm com}<\kappa F_{12}$, where $F_{i,\rm com}$ is the gravitational force between the pair and any neighbor halo $i$ within a $5 \, \hmpc$ radius of the pair center-of-mass,
 $F_{12}$ is the force between the pair, and $\kappa$ is a constant parameter. The isolation criterion becomes increasingly strict for decreasing values of $\kappa$. The Milky Way and M31 do not have massive neighbors within $5\ \mpc$, and  should thus have $\kappa<0.1$. The actual value of $\kappa$ is, however, uncertain, and  we use $\kappa=0.25$ based on our previous tests reported in \citet{2013ApJ...770...96G}.

An additional selection criterion is intended to mimic the absence of massive clusters in the immediate vicinity of the Local Group. We require that halos in the LG sample have no neighbor halo with mass  $M_{200c}>1.5 \times 10^{14}\ \rm M_{\odot}$ within $12$ Mpc. The mass and distance limits are somewhat lower than the actual values 
for the Virgo Cluster \citep[e.g.,][and references therein]{2001AA...375..770F,1995MNRAS.274.1093N} to allow for a larger number of systems. 

We found $4177$ pairs in the snapshot at $z=0$ under these constraints, and for the full composite sample using $10$ more snapshots at $z<0.1$ we found $45844$ pairs\footnote{Some of these pairs are not independent are repeated in other snapshot, but after including the additional constraints, it is extremely unlikely to have repeats. Nevertheless, if any repeats are identified they are removed at this stage.}, which we use as the sample of LG analogues.
We find that $\approx 80\%$ of LG analogues in this sample are gravitationally bound under the two-body approximation.
Note that the environment criteria are very restrictive: from the initial sample of pairs selected only by mass and separation, less than $1\%$ satisfy the environment criteria.

%-------------------------
\section{Mass likelihood}
\label{sec:like}
%-------------------------

To compute the likelihood distribution for the  \mbox{MW-M31} pairs using the sample of the LG analogues we follow the approach of \citet{2011ApJ...743...40B}, but with modifications described in \citet{2013ApJ...770...96G}. Namely, to compute the likelihood we first select a sample of LG-like pairs satisfying a particular combination of constraints. For each combination of contstraints, the mass likelihood, $P(M)$, is then computed as the normalized distribution of the pair mass. Note that this is different from the procedure used by \citet{2011ApJ...743...40B}, because we compute the likelihood for several combined constraints as the direct mass distribution of halo pairs, rather than multiplying the likelihood distributions for individual constraints as was done by \citet{2011ApJ...743...40B}. Thus, in our procedure we do not  assume that the properties used in different constraints are uncorrelated. Any correlation between constraints is included in the total likelihood.

The specific properties we use in our constraints to define LG pair samples and to construct the mass likelihood for different combinations of constraints are as follows.

\begin{figure*}[!htb]
\begin{center}
\includegraphics[width=.7\linewidth,angle=0]{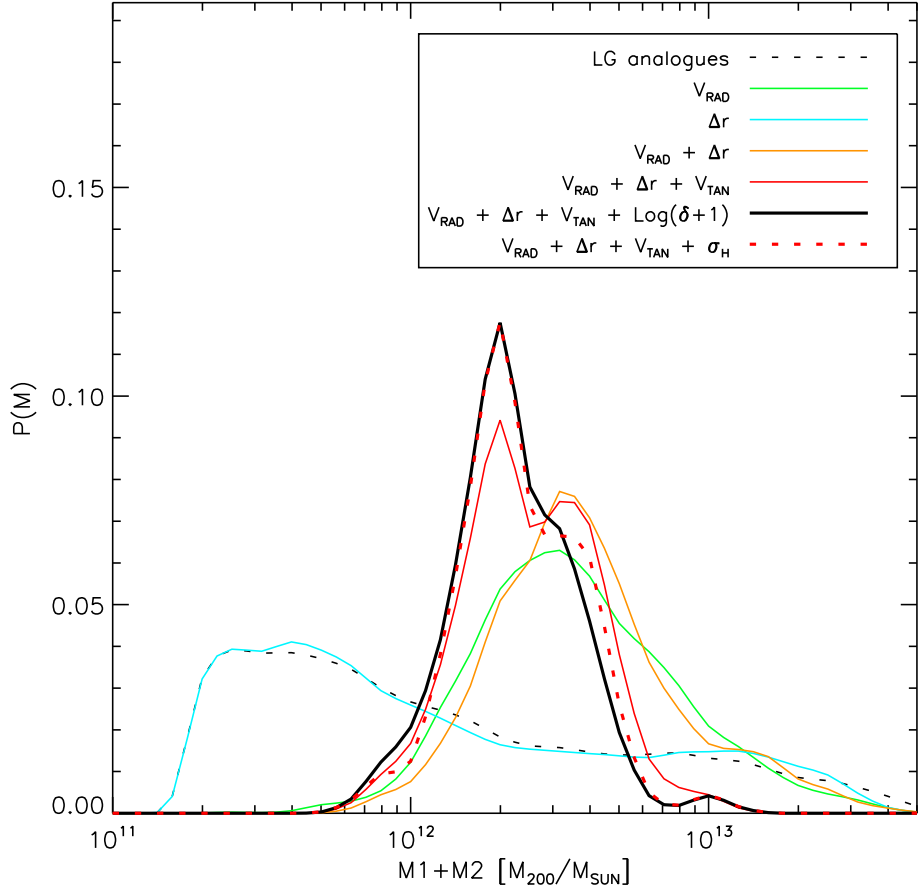}
\end{center}
\caption{
\label{figlgmass1} The likelihood distribution for the sum of $M_{200c}$ masses (mass within radius enclosing density equal to $200c$ times the critical density of the universe) of MW and M31 constructed using the LG halo pair analogues  and using obsevational measurements of the relative separation and motion of MW and M31 with interval corresponding to $2\sigma$ of their measurement errors, as well as constraints on their environment (see legend). The median mass and corresponding 68\% and 90\% confidence intervals for each set of constraints are given in Table 1. } 
\end{figure*}

\begin{itemize}
\item[1.] Galactocentric radial velocity, $V_{\rm RAD}=109.3\pm4.4$ \kms, of M31 measured recently by \citet{2012ApJ...753....7S}. We include in the pairs radial peculiar velocities, the hubble flow at companion distance so we do take into account the Hubble expansion correction.

\item[2.] Distance between M31 and MW, $\Delta r=770\pm40$ \kpc, adopted by \citet{2008ApJ...678..187V} to span the range of recent measurements using different methods, tip of the red giant branch \citep{2001AJ....121.2557D,2005MNRAS.356..979M}, cepheids \citep{2003A&A...402..113J,2004AJ....127.2031K}, and eclipsing binaries \citep{2005ApJ...635L..37R}. 

\item[3.]   The tangential velocity component of M31 relative to the MW is $V_{\rm TAN}=17$ \kms, where the $1\sigma$ upper limit is $V_{\rm TAN}<34.3$ \kms recently derived by \citet{2012ApJ...753....7S}. 

\item[4.] We use the local environment constraint using the local velocity dispersion, $\sigma_{\rm H}$,  and logarithm of the overdensity $\log(1+\delta)$ in a shell with inner radius of $1$ \mpc, and outer radius of $5$ \mpc $\,$ from the pair center-of-mass.
We use constraint of $\sigma_{\rm H}<70$ \kms $\,$to reflect the range of observational estimates (see \S~1).
In the appendix we show that the velocity dispersion is correlated with overdensity, and that our fiducial choice of $\sigma_{\rm H}<70$ \kms\, approximately corresponds to the upper limit on overdensity  $\log(1+\delta)<0.3$.
\end{itemize}

Our fiducial choice for the mass constraints presented in the paper is to use the range of observed values of constraint parameters with interval corresponding to $\pm 2\sigma$ of their observational errors. Namely, we use the following {\it rms} uncertainties:
$\sigma_{V_{\rm RAD}}=4.4$ \kms $\,$for radial velocity and $\sigma_{\Delta r}=40$ \kpc $\,$for separation. 
We do not have a complete information about the confidence interval for the tangential velocity
and we use the estimate of the average tangential velocity of $17$\kms and its 
$1\sigma$ upper limit of $V_{\rm TAN}<34.3$ \kms to extrapolate to $2\sigma$ upper limit of
 $V_{\rm TAN}<51.6$ \kms, and $3\sigma$ of $V_{\rm TAN}<68.6$ \kms.

With these assumptions we select the pairs with corresponding properties $p$ lying in the interval $p\pm 2\sigma_p$. Thus, for example, for the first constraint in Table 1, which uses the combination of radial velocity and pair separation, we select all halo pairs that satisfy isolation criteria described in the previous section and have relative radial velocity $V_{\rm RAD}=109.3\pm 8.8\,\rm km\,s^{-1}$ and $\Delta r=770\pm 80$ kpc. The mass likelihood distribution for the pair sample selected using properties within these intervals is then simply the normalized mass distribution of pairs in the sample. 

We have explored the effect of expanding our sample of pairs by increasing the interval in radial velocity, separation and tangential velocity, we use to select pairs to $p\pm 3\sigma_p$  and found that the mass likelihood is stable. A more detailed description of the tests for different constraints ranges can be found in the Appendix B. 

\begin{table*}[ht]
\begin{center}
\caption{Mass likelihood of MW$+$M31 pairs in LG analogues
}
\label{table1} 
\begin{tabular}{lcccc}
\hline 
\hline
\\
Constraints & $\log(M_{200c}/\msol)$ & $68\%$ conf. internval & $90\%$ conf. interval & $N$ pairs\\
\\
\hline
\\
$V_{\rm RAD} + \Delta r$ &  12.60 &  -0.10 +0.12 & -0.31 +0.45  & 347\\
$V_{\rm RAD} + \Delta r + V_{\rm TAN}$ &  12.45 &  -0.12 +0.11 & -0.25 +0.25 & 88 \\
$V_{\rm RAD} + \Delta r + V_{\rm TAN} + \log(1+\delta)$ &  12.38 &  -0.07 +0.09 & -0.25 +0.24 & 66 \\
$V_{\rm RAD} + \Delta r + V_{\rm TAN} + \sigma_{\rm H}$ &  12.39 &  -0.07 +0.13 & -0.19 +0.27 & 64 \\
$V_{\rm RAD} + \Delta r + V_{\rm TAN} + \log(1+\delta)$ + $1$~Mpc$^a$&  12.62 &  -0.11 +0.13 & -0.28 +0.26 & 66\\
$V_{\rm RAD} + \Delta r + V_{\rm TAN} + \sigma_{\rm H}$ + $1$Mpc&  12.62 &  -0.11 -0.13 & -0.28 +0.27 & 64\\
\\
\hline
%\multicolumn{5}{l}{
%\begin{minipage}{11.7cm}
%{\footnotetext[a]{The last two rows give constraints for the mass enclosed within $1$ Mpc of the barycenter of the pair.}
%\end{minipage}
%\footnotetext[a]{The last two rows give constraints for the mass enclosed within $1$ Mpc of the barycenter of the pair.}
\end{tabular}
\end{center}
\end{table*}

Figure \ref{figlgmass1} shows the likelihood distribution for the sum of $M_{200c}$ masses of the two pair halos obtained for different constraint combinations. In table \ref{table1} we present the corresponding median values, with $68\%$ and $90\%$ confidence intervals and number of pairs in each sample. In addition, we present constraints on the mass within the radius of $1$ Mpc from the pair barycenter in the last two rows of the table. 

Figure \ref{figlgmass1} shows that the radial velocity of M31 provides the main constraint on the masses. Nevertheless, inclusion of the tangential velocity constraint eliminates the tail of objects at very high masses and shifts the peak of the likelihood to lower masses. This is because imposing constraint of low tangential velocity removes more massive pairs with higher orbital energies for a given fixed range of radial velocities. 

The local density and velocity dispersion constraints do not affect the peak of the likelihood distribution but slightly narrow the width of the likelihood.
Overall, we find that inclusion of the environment constraints in the likelihood calculations makes no significant difference: they shift the median and confidence intervals to masses $\approx 15\%$ lower. 

We have explored the mass ratio distribution
\footnote{Defined as the ratio between the halo mass of the smallest and largest pair member. The mass ratio of the MW/M31 pair is quite uncertain, but recent papers point toward a mass ratio close to unity, where M31 is somewhat more massive than the MW \citep{2009MNRAS.393.1265K,2009ApJ...700..137R,2009MNRAS.397.1990B}}
for different samples following constraints from table \ref{table1} and  found no effect of additional constraints on the mass ratio on the mass likelihood. 

We also compute the statistical errors in the shape of the distribution due the low number of pairs in the samples, in figure \ref{figlgmasserror} we show errors for the two most relevant samples: $V_{\rm RAD} + \Delta r + V_{\rm TAN}$ (red), and $V_{\rm RAD} + \Delta r + V_{\rm TAN} + \log(1+\delta)$ (black). We see there is no significant effect on the width of both distribution produced by the statistical errors, however the bimodality shape observed for red curve is more likely to be a fluctuation due errors rather than a real feature.

\begin{figure}[!t]
\includegraphics[width=.95\linewidth,angle=0]{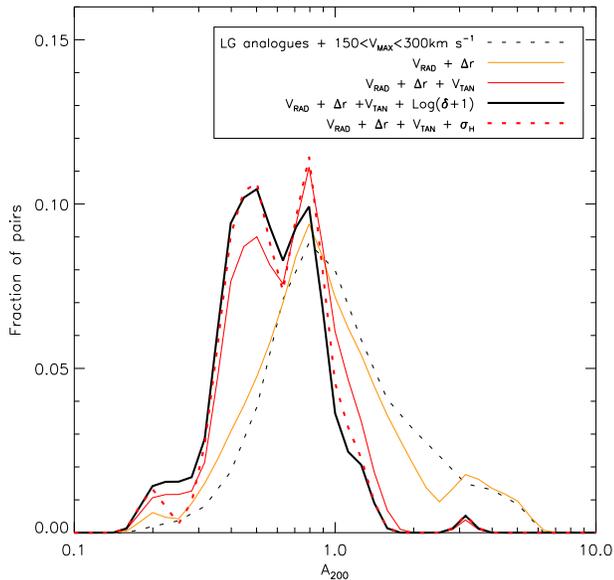}
\caption{
\label{figavalue} Ratio of the true pair mass and the TA mass estimate, $A_{200}$, for different combinations of constraints. In the case of radial velocity and separation constraints (orange), the median $A_{200}$ is close to unity, but when tangential velocity constraint is included, the TA overestimates the true mass pushing the distribution to lower $A_{200}$ values.}
\end{figure}

\section{Comparison with the previous timing argument mass estimates}
\label{sec:ta}

\citet[][hereafter LW08]{2008MNRAS.384.1459L}, computed the bias and error distribution of the TA estimator using LG-like systems in the Millenium Simulation. On average, they found good match between the true $M_{200c}$ masses of MW$+$M31 halos and the $M_{\rm TA}$ masses: $M_{\rm TA}=5.32\pm 0.48 \times 10^{12}$ \msol, or $\log{M_{\rm TA}/M_{\odot}}=[12.68,12.76]$, which not even overlaps with our best $90\%$ confidence interval from table \ref{table1}(Third row).
However, if we compute the TA mass using equations $(1-3)$ from their paper, but with updated radial velocity and separation values, we obtain a somewhat lower value of $M_{\rm TA}=4.14\pm0.60 \times 10^{12}$ \msol ($\log{M_{\rm TA}/M_{\odot}}=[12.55,12.68]$) in good agreement with \citet{2012ApJ...753....8V}. The decrease is due primarily to the lower radial velocity value compared to that used by LW08. Therefore, the TA mass range computed with updated velocity values is in good agreement with our likelihood estimate without taking into account tangential velocity and local environment (first line in Table 1). The somewhat lower estimate in our analysis is due then primarily to the tangential velocity and environment constraints, which shift the peak of the likelihood to smaller masses. 

In figure \ref{figavalue}, we show distribution of the ratio, $A_{200}=M_{200c}({\rm MW+M31})/M_{\rm TA}$ (after LW08), of the true pair mass and the TA mass for different samples of LG analogues\footnote{The TA estimate is computed using age of the universe at $z=0$ even if some pairs are found at $z>0$, but using corrected age for them makes a very little difference and surely the MW and M31 did not start evolving at $a=0$, so using current age of the universe is an overestimate anyway.}. 
The LG analogues sample with an additional cut of $150<V_{\rm max}<300$ \kms (corresponding to the broad $V_{\rm max}$ selection of LW08), shows a distribution with median $A_{200} \approx 1.1$. If we use a narrower $V_{\rm max}$ range, the distribution also becomes narrower, in agreement with results of LW08 (see their Figures 1 and 2). For the radial velocity and distance constraint (orange) the median $A_{200} \approx 0.97$ but with a similar scatter.

Inclusion of the tangential velocity constraint in the selection of pairs (solid red), results in a narrow distribution with the median shifted to smaller value of $A_{200} \approx 0.75$, while inclusion of the additional local density or velocity dispersion constraint shifts the median to $A_{200}\approx 0.62$. This shows explicitly that the TA estimates works quite well for average halo pairs of separations and radial velocity. However, pairs with additional constraints on the tangential velocity and local density have systematically lower masses compared to the TA estimate. In other words, for such pairs the timing argument estimate overestimates mass by a factor of $\approx 1.3-1.6$. This explains the systematic difference between our fiducial constraint from the likelihood, $\log M_{200c}({\rm MW+M31})\approx 12.38^{+0.09}_{-0.07}$ and the TA estimate $\log M_{200c}({\rm MW+M31})\approx 12.62^{+0.06}_{-0.07}$.

There are discrepancies between our results and those of LW08. In particular, LW08 show in their figure $6$ that the median $A_{200}$ does not decrease significantly for $V_{TAN}<86$ \kms. However, we find that $A_{200}$ depends on the  mass distribution of halos in the sample, decreasing with decreasing halo mass. 
In addition, we find that sensitivity of the $A_{200}$ on $V_{TAN}$ constraint also depends on the median mass and the mass range of the sample: for halo samples with narrower mass ranges the effect of the $V_{TAN}$ on $A_{200}$ is weaker.

Specifically, for different mass ranges (masses in units of $M_{200c}/10^{12}\,M_{\odot}$) we measure the median value, $\overline{A}_{200}$:
$M=[0.5,1.0]$ then $\overline{A}_{200}=0.62$;
$M=[1.0,1.5]$ then $\overline{A}_{200}=0.71$; 
$M=[1.75,2.25]$ then $\overline{A}_{200}=0.90$; 
$M=[2.0,3.0]$ then $\overline{A}_{200}=1.06$; 
$M=[3.0,5.0]$ then $\overline{A}_{200}=1.22$.
Thus, we find that $A_{200}$ depends on halo mass of the halos in the pair.
Therefore, because inclusion of $V_{TAN}$ constraint lowers the mass of halos in the sample, we obtain lower value of $\overline{A}_{200}$.
We also made tests for wider ranges of $V_{RAD}$ or $\Delta r$ and find that the wider range does not significantly change the average mass of the underlying mass distribution and median $A_{200}$ values. 

Our LG analogues are allowed in a wide range of masses (see section $3$), and our main sample (row 3, table 1) contains pairs with masses within the range $[1.4,4.4] \times10^{12}\,M_{\odot}$, with an average mass of $\approx 2.4 \times10^{12}\,M_{\odot}$. Given that we do not constraint individual masses of each pair member, the individual halo masses can span a wide mass range from $\sim 0.1$ to $4.4 \times10^{12}\,M_{\odot}$ with an average mass of $\sim 1.2 \times10^{12}\,M_{\odot}$. Thus, they include rather small halos. In contrast, LW08 halo masses lie in the range $[0.8,7] \times10^{12}\,M_{\odot}$ for each pair member, with average halo mass of $\sim 2.5 \times10^{12}\,M_{\odot}$\footnote{We convert LW08 $V_{MAX}$ ranges to $M_{200C}$ by computing the relation directly from the simulation.}. 
So the LW08 sample has narrower mass range and a higher average mass, which is the reason they find a weak sensitivity of $A_{200}$ to the $V_{TAN}$ constraint. For the mass range and average mass of the LW08 sample, our test results  for halo sample with average mass of $\sim 2.5 \times10^{12}\,M_{\odot}$, gives value of $\overline{A}_{200}=1.06$ quite consistent with LW08.

It is worth noting that it is quite surprising that the TA estimate works to within a factor of two, given how idealized the model underlying such estimate is. For example, the MW and M31 are approximated as point masses of constant mass on a purely radial orbit and surrounding mass distribution is neglected. At the same time, the mass evolution of MW and M31 is neglected as well. Finally the evolution of MW and M31 is envisioned within expanding background corresponding to the mean density of the universe and thus any dependence of expansion on the local overdensity is neglected. Given the simplicity of the model and a number of assumptions, it is quite remarkable that this model provides a reasonable ballpark estimate of mass. However, sensitivity to the tangential velocity and environment that we find shows that the accuracy of the TA estimate is ultimately limited. 

Finally, we note that the mass estimate we derive from the likelihood is  in reasonably good agreement with the recent abundance matching results \citep[][see their Appendix]{kravtsov_etal14}, which for the stellar masses of $5\times 10^{10}\rm \ M_{\odot}$ and $9\times 10^{10}\rm \ M_{\odot}$ for the MW and M31, respectively, indicate average halo mass $M_{200}({\rm MW+M31})\approx 4.3\times 10^{12}\rm\ M_{\odot}$ (or $\log_{10}M_{200c}({\rm MW+M31})\approx 12.6$). This abundance matching result is based on the new measurement of the stellar mass function by \citet{bernardi_etal13}, which corrects significant photometric errors in the standard SDSS magnitudes. The average scatter around this average value is thought to be $\approx 0.2$ dex and although the scatter is large the agreement is encouraging, especially because previous abundance matching results by \citet{moster_etal13} and \citet{behroozi_etal13a}, based on older estimates of the stellar mass functions with SDSS photometry, indicated very large average mass of $M_{200c}({\rm MW+M31})\approx 8-10\times 10^{12}\rm\ M_{\odot}$ for the stellar masses of MW and M31. Reconciling low mass of the Local Group with abundance matching results would require assumption that MW and/or M31 are outliers from the average $M_{\ast}-M_{200c}$ relation. However, better agreement with the new abundance matching $M_{\ast}-M_{200c}$ based on the stellar mass function of \citet{bernardi_etal13} indicates that halo masses of the MW and M31 are consistent with the masses expected from the mean $M_{\ast}-M_{200c}$ relation. 

\section{Discussion and conclusions}
\label{sec:conc}

We define the LG analogues in the Bolshoi simulation of {\LCDM} cosmology and estimate the MW-M31 pair mass likelihood in such systems. The analogues are selected  as halo pairs using broad criteria. The sample is then used to estimate likelihood distribution of mass using several observed properties of the actual Local Group, namely separation, radial and tangential velocity, and density of the local environment. To characterize the latter, we compute the DM overdensity and particle velocity dispersion, $\sigma_{\rm H}$, within $5$ \rm{Mpc} from halo pair center of mass. We found a tight correlation between local overdensity and velocity dispersion estimated within 5 Mpc (see Appendix), so that constraint on overdensity is approximately equivalent to the constraint on the velocity dispersion. To set the environment constraint we require  $\sigma_{\rm H}<70$ \kms\, (or $\log{(\delta+1)}<0.3$) based on the observational values of this dispersion reported in the literature. 
At a given snapshot around $z=0$, about $2\%$ of the MW-sized ($M_{200}\sim10^{12}$ \msol) halos satisfy our broad LG analogue criteria, and less than $5\%$ of these two per cent satisfy the additional orbital and environment constraints.

We have shown that the main parameter controlling the MW-M31 mass likelihood is radial velocity. However, we also show that the likelihood is sensitive to the constraints on tangential velocity. In particular, we find that mass likelihood peak shifts to lower masses when constraint on the tangential velocity is included. 
This is because this constraint eliminates massive pairs with a given radial velocity range. 
Note that there is no such sensitivity to the tangential velocity in the timing argument estimate because neither the tangential velocity nor environment are taken into account in such estimate. Indeed, the mass constrain we derive from the likelihood is in good agreement with the mass estimate from the timing argument when only radial velocity and separation are used as constraints:  $M_{\rm TA}=4.14\pm0.60 \times 10^{12}$ \msol, in agreement with \citet{2012ApJ...753....8V} results, but somewhat lower than LW08 due to lower updated value of the radial velocity used in our estimate. However, when we add tangential velocity and environment constraints the median of the likelihood shifts to lower masses by a factor of $\approx 1.6$.
We show explicitly that for pairs with low tangential velocities and low local overdensity and velocity dispersion, the timing argument overestimates true masses of the pair by an average factor of $\approx 1.6$, thereby explaining the lower values derived from the likelihood method. These values are summarized in Table \ref{table1}: our fiducial mass estimate obtained including the local density constraint is $M_{\rm MW,200c}+M_{\rm M31,200c}=2.40_{-0.36}^{+0.55}\times10^{12}M_{\odot}$ ($68\%$ confidence interval).
For this sample we have also computed the DM mass enclosed within $1$ \rm{Mpc} from the pair center of mass: $M_{\rm LG}(r<{\rm 1Mpc})=4.17_{-0.93}^{+1.45}\times10^{12}M_{\odot}$ ($68\%$ confidence interval). 

Overall, the values we deduce for the sum of the Milky Way and M31 halo masses are consistent with existing constraints on the individual halo masses of the Milky Way \citep[see, e.g.,][and references therein]{boylan_kolchin_etal12,boylan_kolchin_etal13} and M31 \citep{widrow_dubinsky05}. Given that the estimates of halo mass using satellite velocity \citep[e.g.,][]{boylan_kolchin_etal13} can give robust lower limits to individual halo masses, combination of the estimates of the combined LG mass and individual masses should help to narrow down the range of possible masses for our Galaxy and for our closest neighbor, M31.

\acknowledgments 
We would like to thank Anatoly Klypin for making the Bolshoi simulation and the BDM halo catalogs publicly available. 
This work was supported by NSF via grant OCI-0904482.
AK was in addition supported in part by NSF
grants AST-0807444 and by the
Kavli Institute for Cosmological Physics at the University
of Chicago through the NSF grant PHY-0551142 and PHY-
1125897 and an endowment from the Kavli Foundation. We have made extensive use of the NASA Astrophysics Data System
and {\tt arXiv.org} preprint server.

\appendix

\section{Cold Local Hubble Flow}

In order to characterize the local environment of the LG we use the velocity dispersion $\sigma_{\rm H}$ of nearby galaxies. It is known that this velocity dispersion is rather low within a few Mpc from the Local Group, compared to the velocity dispersion expected for MW-sized halos in simulations \citep[e.g.,][]{1997NewA....2...91G}.
This ``coldness'' of the local Hubble flow was noted for quite some time in studies measuring the Hubble constant with local galaxies \citep{de_vaucouleurs_etal58, 1975ApJ...196..313S}. 
The values of velocity dispersion were consistently found to be around $\sigma_{\rm H} \sim 60$ \kms $\,$ up to $8$ \rm{Mpc} \citep{1975ApJ...196..313S,1986A&A...170....1G,2001A&A...368L..17E,2009MNRAS.393.1265K,2003AA...398..479K,2005MNRAS.359..941M,2009MNRAS.395.1915T,2003ApJ...596...19K,2011MNRAS.415L..16A}.
However, the local velocity dispersion increases with the maximum radius adopted to measure it \citep{1975ApJ...196..313S,2005MNRAS.359..941M,2009MNRAS.395.1915T}, and thus the specific value of the velocity dispersion used for constraints should correspond to the radius used in observations.

\begin{figure}[tb]
\includegraphics[width=.95\linewidth,angle=0]{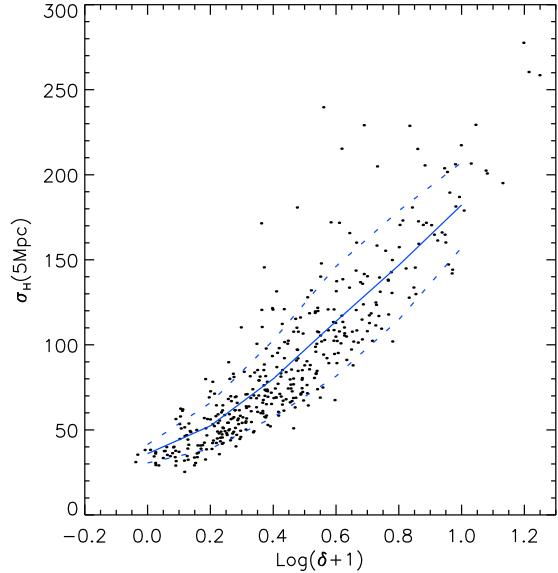}
\caption{
\label{figflow1} The local overdensity and velocity dispersion relation for the MW-sized halos in the Local Group analogues sample. Blue lines show the mean and standard deviation of the distribution. Density and velocity dispersion are computed using DM particles within a shell ranging from $1$ to $5$ \mpc.}
\end{figure}

Furthermore, the methodology to compute the local velocity dispersion must also be taken into account when comparing different results; i.e., \citet{2003AA...398..479K} found $\sigma_{\rm H}=85$ \kms $\,$ within $5$ \mpc, but the estimate drops to $\sigma_{\rm H}=41$ \kms $\,$ when members of the M81 and Cen A groups are removed.
We adopt the conservative value of $\sigma_{\rm H} < 70$ \kms $\,$ as a constraint for the mass likelihood computation. 
The coldness of the local flow can be attributed to the relative isolation of the LG, and the relatively low density of the LG environment \citep{2003ApJ...596...19K,2009MNRAS.397.2070M}. 

\citet{2005MNRAS.359..941M} found correlation between  $\sigma_{\rm H}$ and local density in \LCDM numerical simulations and pointed out that observational velocity dispersion measurements around the Local Group imply overdensity of $-0.1<\delta \rho/\rho<0.6$ on the scale of $7$ \rm{Mpc}.
We explore the $\sigma_{\rm H}$--overdensity relation for our sample of LG analogues, estimating $\sigma_{\rm H}$ using halos and dark matter particles with respect to the pair center of mass.  For the former we compute the radial peculiar velocity of host halos with maximum circular velocities of $V_{\rm max}>50$ \kms. For the latter estimate we use DM particles outside $3 \times R_{200}$ from any halos with $V_{\rm max}>50$.

Halos should be closer to galaxy tracers used in observations, but for some pairs the number of halos is too low for an adequate $\sigma_{\rm H}$ computation. On the other hand, using particles gives a robust $\sigma_{\rm H}$ computation in all cases, but makes comparison with observations more ambiguous.
There is an overall scatter of $\sim 40$ \kms $\,$ between the two estimates of $\sigma_{\rm H}$, which decreases to $\approx 25$ \kms, for velocity dispersions lower than $100$ \kms.
In both cases, $\sigma_{\rm H}$ is computed in shells with the radius ranging within $1-5$ \mpc. The lower limit is set because the Hubble flow is observed only at $R_0 \gtrsim 1$ \rm{Mpc} \citep[e.g.,][]{karachentsev12}. This can be used to put an upper limit to the mass of the LG \citep{2001A&A...368L..17E,2002AA...389..812K,2009MNRAS.393.1265K}.

In figure \ref{figflow1}, we show the $\sigma_{\rm H}$--overdensity relation estimated around the LG analogues, in which   each pair member is  in the mass range of $0.8-2.9 \times 10^{12}$ \msol $\,$chosen to follow sample definition from \citet{2013ApJ...770...96G}.
We see a fairly tight relation at low $\sigma_H$ values in agreement with \citet{2005MNRAS.359..941M} with $20\%$, $43\%$, and $65\%$ of the LG analogues having $\sigma_H$ lower than $50$, $70$, and $100$ \kms, which corresponds to the average overdensities of $\log{(1+\delta)}=0.156$, $0.237$, and $0.303$, respectively.
For the mass likelihood estimate, we use the constraint $\log(1+\delta)<0.3$ corresponding to systems with $\sigma_{\rm H} < 70$ \kms $\,$on average.

\section{Sensitivity of the mass likelihood distribution to constraint choices}

The pair separation, radial and tangential velocity are the main orbital constraints used for the mass likelihood computation.
They are very restrictive due their small associated errors resulting in small number of halo pairs. The samples can be increased if we relax these constraints and assume instead that the errors for a particular parameter are two or three times larger than the actual errors.  
In this section we investigate the effect of such choices used as a constraint in calculation of the likelihood distribution. Larger adopted error allows to increase the size of the halo samples and decrease the associated Poisson errors. However, it means that we allow for inclusion of  objects less consistent with observational constraints. The actual choice of the error is
a trade-off between these two considerations. 

The $1\sigma$ error for distance and radial velocity are $\sigma(\Delta r)=40$ \kpc, and $\sigma(V_{\rm RAD})=4.4$ \kms. The mean value of the tangential velocity is $17$ \kms with $1\sigma$ upper limit of $V_{\rm TAN}<34.3$ \kms, which we extrapolate to $2\sigma$ and $3\sigma$ upper limits of $V_{\rm TAN}<51.6$ \kms $\,$and $V_{\rm TAN}<68.6$ \kms, respectively.
We repeat our likelihood calculations using errors of constraint parameter inflated by a factor of three for each constraint separately and for all combined constraints. Here we do not 
include environment constraint, given that we found that their effect is relatively small.

In the first test, we keep the errors for distance and tangential velocity fixed to their respective $1\sigma$ values, while we change the error of $\sigma(V_{\rm RAD})$.
Results are listed in Table \ref{tablea1}, which shows that the median mass value increases when the error of radial velocity is increased  from $\pm 4.4$ \kms to $8.8$ \kms, but is not sensitive to further increase due to combined constraints to all of the other parameters. The $68\%$ and $90\%$ confidence intervals also increase somewhat with increasing error. 
The number of pairs increase from $12$ to $60$ for $1\sigma$ to $4\sigma$ values.

\begin{table}[t]
\begin{center}
\caption{Mass likelihood dependence on $V_{\rm RAD}$ constraint amplitude}
\label{tablea1}
\begin{tabular}{lccc}
\hline
\hline
\\
$\sigma(V_{\rm RAD})/(4.4$ \kms$)$& $\log(M_{200}/$\msol$)$ & $90\%$ c. i. & $68\%$ c. i.\\
\\
\hline
\\
1.0 & 12.38 & -0.13  +0.29 & -0.05  +0.19 \\
1.5 & 12.35 & -0.29  +0.32 & -0.10  +0.17 \\
2.0 & 12.51 & -0.31  +0.27 & -0.19  +0.11 \\
2.5 & 12.48 & -0.28  +0.22 & -0.17  +0.12 \\
3.0 & 12.51 & -0.31  +0.19 & -0.17  +0.09 \\
3.5 & 12.51 & -0.32  +0.27 & -0.17  +0.09 \\
4.0 & 12.51 & -0.40  +0.28 & -0.17  +0.11 \\
\\
\hline

\end{tabular}
\end{center}
\end{table}

In the second test, we have kept the errors of radial and tangential velocity fixed at $1\sigma$, but increased error of distance, $\sigma (\Delta r)$ (Table \ref{tablea2}).
The median mass and $68\%$ confidence interval are not sensitive to increases in the distance errors, while $90\%$ confidence interval increases somewhat. At $3\sigma$ the number of pairs increases to $40$.

\begin{table}[t]
\begin{center}
\caption{Mass likelihood dependence on $\Delta r$ constraint amplitude}
\label{tablea2}
\begin{tabular}{lccc}
\hline
\hline
\\
$\sigma (\Delta r)/(40$ \kpc$)$& $\log(M_{200}/$\msol$)$ & $90\%$ c. i. & $68\%$ c. i.\\
\\
\hline
\\
2.0 & 12.36 & -0.12 +0.31 & -0.04 +0.17 \\
3.0 & 12.37 & -0.24 +0.30 & -0.07 +0.13 \\
\\
\hline
\end{tabular}
\end{center}
\end{table}

In the third test we have kept the radial  velocity and separation errors fixed at $1\sigma$, while varying the adopted error of $\sigma(V_{\rm TAN})$ (Table \ref{tablea3}). Increase of the tangential velocity constraint to $51.6$ \kms ($2\sigma$) does not change the median, while increase to $68.6$ \kms ($3\sigma$) leads to increase of the median mass value and the confidence intervals. The number of pairs increases to $20$ and $31$ for $2\sigma$ and $3\sigma$ respectively.

\begin{table}[t]
\begin{center}
\caption{Mass likelihood dependence on $V_{\rm TAN}$ constraint amplitude}
\label{tablea3}
\begin{tabular}{lccc}
\hline
\hline
\\
$V_{\rm TAN}/($\kms $)$ & $\log(M_{200}/\msol)$ & $90\%$ c. i. & $68\%$ c. i.\\
\\
\hline
\\
51.6 & 12.38 & -0.14 +0.39 & -0.07 +0.20 \\
68.6 & 12.54 & -0.29 +0.23 & -0.19 +0.05 \\
\\
\hline
\end{tabular}
\end{center}
\end{table}

Finally, when we vary errors of all of the constraining parameters simultaneously (Table \ref{tablea4}), the median mass value increases by less than $1\sigma$ when errors are inflated by a factor of two. The main effect on the confidence intervals is to increase the range of mass values smaller than the median, while the upper error bar does not change and even decreases slightly.

\begin{table}[ht]
\begin{center}
\caption{Mass likelihood dependence on increasing all constraints amplitudes}
\label{tablea4}
\begin{tabular}{lccc}
\hline
\hline
\\
$\sigma({\rm test})/\sigma$ & $\log(M_{200}/\msol)$ & $90\%$ c. i. & $68\%$ c. i.\\
\\
\hline
\\
1.0 & 12.38 & -0.13 +0.29 & -0.05 +0.19 \\
2.0 & 12.45 & -0.25 +0.25 & -0.12 +0.11 \\
3.0 & 12.44 & -0.32 +0.27 & -0.13 +0.11 \\
\\
\hline
\end{tabular}
\end{center}
\end{table}

These tests indicate the our results for the mass likelihood do not depend sensitively to our fiducial choice to inflate observed errors by a factor of two. In fact, if anything, the derived mass constraint for the actual observational errors is slightly smaller with smaller error bars. In this case, however, the sample contains only $12$ pairs. 
We therefore think  that our fiducial choice is more conservative.

\begin{figure}[!htb]
\begin{center}
\includegraphics[width=.95\linewidth,angle=0]{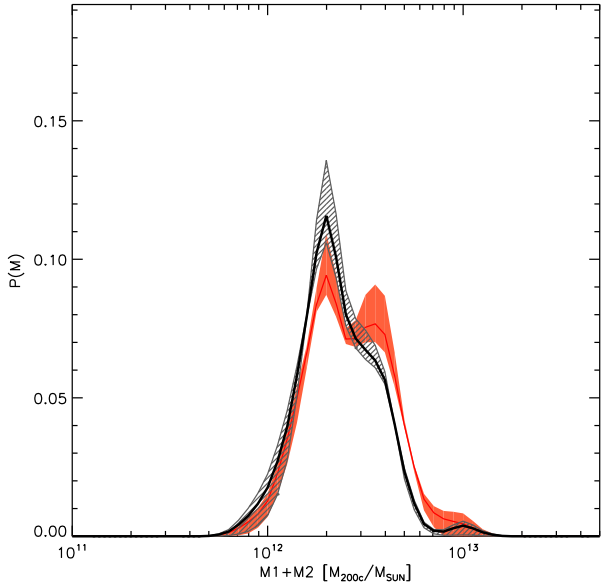}
\end{center}
\caption{
\label{figlgmasserror} The likelihood distribution for the sum of $M_{200}$ masses, similar to figure \ref{figlgmass1} for the $V_{\rm RAD} + \Delta r + V_{\rm TAN}$ (red), and $V_{\rm RAD} + \Delta r + V_{\rm TAN} + \log(1+\delta)$ (black) samples. Error contours computed using jacknife estimator are included. There is no significant effect on the width of both distribution produced by the statistical errors coming from samples size.  }
\end{figure}

\bibliography{satbib}

\end{document}